
\input phyzzx
\overfullrule=0pt
\def\n{\noindent}
\def\p{\partial}

\def\gl{\lambda^2}

\def\sk{\sqrt{\kappa} }
\def\IN{\relax{\rm I\kern-.18em N}}

\null
\rightline {UTTG-32-93}
\rightline {December 1993}

\title{On Black Hole Singularities in Quantum Gravity
\foot{Work supported in part by NSF grant PHY 9009850 and
R.~A.~Welch Foundation.}}

\author{Jorge G. Russo }
\address {Theory Group, Department of Physics, University of
Texas\break
Austin, TX 78712}

\singlespace
\abstract
We show that absence of space-like boundaries in 1+1 dimensional
dilaton gravity implies a catastrophic event at the end point of
black hole evaporation. The proof is completely independent of the
physics at Planck scales, which suggests that the same
will occur in any theory of quantum gravity which
only admits trivial space-time topologies.

\bigskip
PACS Categories: 04.60+.n, 12.25.+e, 97.60.Lf
\endpage

\Ref\hape{R. Penrose, Phys. Rev. Lett. 14 (1965) 57;
S.~W.~Hawking, Proc. R. Soc. London A294 (1966) 511;
A295 (1966) 460; A300 (1967) 187;
S.~W.~Hawking and R. Penrose, Proc. R. Soc. London A314 (1970) 529.}

\Ref\hawi{S.~W.~Hawking,
Comm. Math. Phys.  43 (1975) 199; Phys. Rev.  D14 (1976) 2460.}

\Ref\yo{ J.G. Russo, {\it Quantum black holes: non-perturbative corrections
and no-veil scenario},  preprint UTTG-22-93 (1993).}

\Ref\cghs{C.~G.~Callan, S.~B.~Giddings, J.~A.~Harvey and
A.~Strominger, Phys. Rev.  D45 (1992) R1005.}

\Ref\dpage{J.A.~Harvey and A.~Strominger, {\it Quantum aspects of black holes},
 preprint EFI-92-41; S.B.~Giddings, preprint UCSBTH-92-36;
D.N. Page, University of Alberta preprint, Alberta-Thy-23-93,
(1993).}

\Ref\endpoint{J.~G.~Russo, L.~Susskind and L.~Thorlacius,
 Phys. Rev. D46 (1992) 3444; Phys. Rev. D47 (1993) 533. }

\Ref\gidnel{S.~B.~Giddings and W. Nelson, Phys.Rev. D46 (1992) 2486.}


For some time it was widely believed that the classical theory
of general relativity does not predict singularities, and that they were
just an artifact of the high degree of symmetry of spherical solutions.
A slight perturbation to spherical collapse would produce nonsingular
``bouncing" solutions, as occurs in the Newtonian theory of gravity.
This idea was abandoned after the appearance of the singularity theorems
by Hawking and Penrose [\hape]. They showed that a curvature singularity
in the circumstance of gravitational collapse is inevitable provided
the energy-momentum tensor of matter satisfies a positivity
condition.
The hope that remained is that quantum effects might prevent
the singularity from occurring or might smear it out in some way,
and the problem would be absent in a full
quantum treatment.

In the case of black hole formation, the singularity predicted by
classical gravity represents a space-like boundary of the Universe.
The standard picture of black hole formation and evaporation  [\hawi ]
is described by a Penrose diagram as indicated in fig. 1. If no
curvature singularity is actually present in the full quantum theory,
it is  plausible that the topology will simply be equivalent to that
of Minkowski space-time, as shown in fig. 2, since
nothing would prevent matter to propagate towards the future null infinity.
This is the basic premise of geodesic completeness.  If the space-like
boundary persisted there would be geodesics with finite affine lenght
in one direction and the space-time would be geodesically incomplete,
which entails a serious physical pathology.
In fig. 2 there is no global horizon and, in principle, a unitary evolution
described by an $S$-matrix is expected. A recent study of this
scenario in the context of dilaton-gravity is given in ref. [\yo ], where
it is implemented through non-perturbative corrections.
Let us note that it is not impossible that
 there is a space-like boundary and nevertheless no curvature singularity.
Near strong curvature regions, matter might stop propagating because, e.g.,
the theory becomes ``topological". This would still be regarded as a
singular space-time according to standard definitions in general relativity.
Other variants are, for example, that the space-time boundary
effectively subsists because there is a whormhole connecting to
other world, or there is multiple formation of baby universes, etc.
As far as the information problem is concerned, all these scenarios
are equivalent to the case with  curvature singularity of fig. 1,
inasmuch as there {\it is} a
problem of information loss; in the case of fig. 2 information can simply
reappear by reflection on the boundary [\yo ] (we shall return on this
later).

We would like to explore the consequences of starting from
the following postulates:

\noindent 1) {\it The final state after black hole evaporation is
the vacuum }(``$CPT$ invariance" condition).

\noindent 2) {\it The space-time topology for black hole evaporation
is that of Minkowski space-time } (absence of ``singularities" in full
quantum theory).

\n Most quantum field theorists will find these postulates to be
a {\it sine qua non}.
In particular, they provide a natural
framework to define a unitary S-matrix and maintain quantum coherence.
However, we shall see that, under these assumptions, a catastrophic event
at the end point of Hawking evaporation is inevitable. It is an
outgoing shock-wave propagating at the speed of light and
carrying Planck scalar curvature. This is a smoother version of what was
referred by Hawking as a ``thunderbolt singularity" (a singularity which
spreads out to infinity on a space-like or null path).

Let us consider quantum gravity in 1+1 dimensions.
A simplified model for black hole formation and
evaporation, known as the CGHS model, was introduced in ref. [\cghs ].
This model permits to study the Hawking phenomenon in detail including
back reaction, avoiding all the mathematical complications of
higher-dimensional
theories (for reviews and references therein see ref. [\dpage ]).
Let us consider the model introduced in ref. [\endpoint ].
The effective action which includes the anomaly term can be written
as
$$
S={1\over 2\pi}\int d^2x\sqrt {-g}\big[ e^{-2\phi}\big( R+4
(\nabla\phi)^2+4\lambda^2\big)-{1\over 2}\sum_{i=1}^N (\nabla f_i)^2
-{\kappa\over 4} \big( R{1\over \nabla ^2}R +2\phi R\big)\big]\ ,
\eqn\accion
$$
where $f_i$ are $N$ conformal fields and $\kappa=(N-24)/12 >0$.
For the purposes of this paper, the classical part is actually
suffices, but we will often keep track of the first quantum corrections
for clarity in the derivation.
In the conformal
gauge $g_{++}=g_{--}=0$, $g_{+-}=-{1\over 2} e^{2\rho}$, the effective
action containing the conformal anomaly can be written as
$$S={1\over \pi}\int d^2x \big[ -\p_+\chi \p_-\chi +
\p_+\Omega \p_-\Omega +\gl e^{(2/\sqrt{\kappa })(\chi-\Omega )}
+{1\over 2}\sum_{i=1}^N\p_+f_i\p_-f_i\big]\ ,
\eqn\action
$$
where $\Omega={\sqrt{\kappa}\over 2}\phi +{e^{-2\phi}\over\sqrt\kappa } ,
\ \chi -\Omega =\sk (\rho-\phi ) .$
The constraints are
$$
\kappa t_{\pm}(x^\pm )=-\p_\pm\chi \p_\pm\chi + \p_\pm\Omega \p_\pm\Omega
+\sqrt{\kappa}\p^2_\pm\chi +{1\over 2}\sum_{i=1}^N\p_\pm f_i\p_\pm f_i\ .
\eqn\constraints
$$
The functions $t_\pm (x^\pm )$ reflect the non-local nature of the anomaly
and are determined by boundary conditions.
The solution to the semi-classical equations
of motion and the constraints,
for general distributions of incoming matter, is given  in Kruskal
coordinates by
$$
\Omega=\chi=-{\gl\over\sk }x^+\big( x^-+{1\over \gl }P_+(x^+)\big)
+{M(x^+)\over \sk\lambda } -{\sk \over 4}\ln (-\gl x^+x^-)\ ,
\eqn\solutions
$$
where $M(x^+)$ and $P_+(x^+)$ respectively represent total energy and
Kruskal momentum of the incoming matter
at advanced time $x^+$:
$M(x^+)=\lambda \int _0^{x^+} dx^+ x^+ T_{++}(x^+)\ ,
\ \ P_+(x^+)= \int _0^{x^+} dx^+ T_{++}(x^+)\  .
$
In the case $T_{++}=0$ one obtains the familiar linear dilaton vacuum,
$e^{-2\phi}=e^{-2\rho}=-\gl x^+x^-$.

Let us assume that originally the geometry is the linear dilaton vacuum
and at some time, which we arbitrarily set at $x^+=1/\lambda $,
the incoming flux is turned on.
As observed in ref. [\endpoint ],
there are two different regimes, according to
whether the incoming matter energy-momentum tensor is less or
greater than a critical flux
$T_{++}^{\rm cr}(x^+)=\kappa /(4 {x^+}{}^2)$ .
In the supercritical regime the boundary line $\phi=\phi_{\rm cr}$
is space-like and one has a time-depending geometry representing the
process of formation and evaporation of a black hole. At the null line
corresponding to the end point
of black hole evaporation, $x^-=x^-_e, \ x^+>x^+_e$,
it is possible to match the solution continuously with the linear dilaton
vacuum.
In the subcritical regime the boundary is time-like and
one needs boundary conditions in order to determine the evolution in the
region in causal contact with the time-like boundary.
It turns out that there are natural, reflecting-type boundary conditions
which uniquely determine the evolution and implement cosmic
censorship hypothesis [\endpoint ].

In weak coupling regions $e^{2\phi }$
the curvature scalar, $R=8e^{-2\rho}\p_+\p_-\rho$, is given by
$$
R=16 e^{-2\phi }\p_+\phi \p_-\phi + 4\gl \ .
\eqn\curvature
$$

Now let us consider the case of formation of a macroscopic black hole
and make use of postulate 2. The Kruskal diagram can be
divided in several regions, as shown in fig. 3. The boundary of the trapped
region is the apparent horizon defined as the locus of  $\p_+\phi\p_-\phi $
(see ref. [\yo ]).
We would like to prove that asymptotic observers at ${\cal J}^+$ will
necessarily experience planckian physics before
entering into the vacuum region V (and thus they will not survive
to report the result of their experiment).
Region II can be defined as the region where the coupling $e^{2\phi}$ is
smaller than, say, $10^{-10}$, which is not in the causal future
of regions with higher couplings, so that the semi-classical solution is
indeed reliable in this region.
At the dividing line $x^-=x^-_0$ the coupling takes
the value $10^{-10}$ at some $x^+$ (which for black holes sufficiently large
it will have to be near $x^+_e$).
For a given $x^-<x^-_0$ and $x^+$ large the solution approaches
to the classical form
$$
e^{-2\rho}=e^{-2\phi}=-\gl x^+\big( x^-+p\big) +{m\over\lambda }\ ,
\eqn\clasica
$$
where $m$ and $p$ are the asymptotic values of $M(x^+)$ and $P_+(x^+)/\gl$.
Region II is in causal contact with strong coupling region so
the solution \clasica\ no longer applies. There are two possibilities:
either the solution is discontinuous at some $x^-_1$ in region II, which would
imply infinite (Planck) curvature in this region (see below),
or it is a continuous function of $x^-$.
We proceed {\it per absurdum}.
We must assume that the coupling $e^{2\phi }$ is
small for every $x^-$ at ${\cal J}^+$. This is for the following reason.
In virtue of postulate 1, at late retarded times $x^-$ the solution
approaches to the vacuum. If there were a region of strong coupling in
between, this would already signify a catastrophic event. Since we are
proceeding per absurdum, we assume that no catastrophe occurs for
asymptotic observers and thus $e^{2\phi }$ must be
small for every $x^-$ at ${\cal J}^+$.
This means that for large $x^+$, i.e. $x^+>>x^+_e$, we can apply the
semi-classical equations of motion, the most general solution being
$$
e^{-2\rho}=e^{-2\phi}=-\gl x^+\big( x^-+p\big) +{m\over\lambda }
+h_+(x^+) +h_-(x^-)\ ,
\eqn\clasii
$$
where $h_\pm(x^\pm)$ are two arbitrary functions to be determined by
additional conditions. But continuity requires that $h_+(x^+)=0$
as $x^+\to \infty$.
Now, it is clear that if the solution has the form
\clasii\  with $h_+(x^+)=0$, at $x^-=-p+\epsilon $, $ \epsilon>0$, the coupling
$e^{2\phi }$ will become strong at large $x^+$.
But this is contrary to our hypothesis. Therefore
we conclude that either $e^{2 \phi} $ must have a discontinuity at some point
in region II, or it must be strong near $x^-\cong -p$. Both cases
represent a catastrophic event. In the first case,
since, for a given $x^+$, $\p_+\phi $ and $e^{-2\phi }$ take finite values
on all $x^-$ in region II, and there exists a point $x^-_1$ where
$\p_-\phi=\infty$, from the expression for the scalar curvature we learn
that $R=\infty$ in the semi-classical approximation, which means that
the curvature must assume at least planckian values.


There are a number of issues which arise some concern.
{}From a physical point of view it seems weird that a
so small planckian object, weighting less than $10^{-5}$ grams,
can be capable of generating  an apocalyptic event as such.
The second question is whether one can actually have a discontinuity
(even if smoothed out at Planck scales) of this sort, and why
it would occur. These two points are related.
The necessary discontinuity must be of the form
$$
h_+(x^+)=\gl x^+(p-v)\ ,
\eqn\disco
$$
where $v$ is not much greater than the Planck lenght, $v<O(1/\lambda )$
(it may be zero or even negative). Such a discontinuity at $x^-_1\sim x^-_e$
can occur if, in the strong coupling region, there is an ingoing
flux of negative energy; for example, an imploding shock wave at the endpoint
of the trapped region which forms simultaneously with the outgoing thunderbolt.
This statement is solely based on our classical intuition, and any quantitative
detail shall depend upon the unknown physics of the Planck scale.
At this stage, we could well consider the discussion on the ingoing
flux finished.
Nevertheless, let us proceed, but aware of the limitations of the
semiclassical approach, and explicitly evaluate
the strenght of this ingoing wave by using the constraint equations.
Disregarding the quantum corrections,  one can integrate eq. \constraints \
and obtain
$$
h_+'(x^+)=-{1\over 2}\sum_{i=1}^N \int _{1\over\lambda }^{x^+} dx^+
\p_+f_i\p_+f_i\ ,
\eqn\implo
$$
or $\int _{1\over\lambda }^{x^+} dx^+ T_{++}(x^+)=-\gl (p-v)$.
This is of the same magnitude of the original incoming Kruskal momentum.
Therefore the discontinuity seems to be due to an ingoing flux of negative
energy whose absolute value is of the same order of the original black hole
mass. Clearly, this is a very rough estimate, since near $x^+_e, x^-_e$
the solution will undoubtedly be very different from $\clasii $.
The form of $T_{++}(x^+)$ depends on the details of the transition from the
trapped region to $x^+>x^+_e$.
In particular, an ingoing shock wave emitted  at $x^+_e ,x^-_e$ would
correspond to having $h_+(x^+)\cong x^+(p-v) \theta (x^+-x^+_e) , \
x^-> x^-_e $ (the fact that this type of dependence on $x^-$
subsists even in the weak coupling region $x^+>>x^+_e$ is  because region
II is in the causal future of a strong coupling region, $x^-\cong x^-_e$).
Alternatively, the transition might be smooth, with a continuous
ingoing flux of negative energy being emanated from the strong coupling region
near the inner apparent horizon, where $h_\pm $ should be replaced
by a more complicated structure (some toy models can be found in
ref. [\yo ]).

The outgoing thunderbolt is analogous to
a bubble wall separating a false vacuum, given by
$e^{-2\phi}=-\gl x^+(x^-+p)$, from the true final vacuum given
by $e^{-2\phi}=-\gl x^+(x^-+v)$.
In the first case the Fock space vacuum $|0\rangle _{{\rm out},p}$
is defined as the state which
is annihilated by the negative frequency modes with respect to
plane waves constructed in terms of the coordinate
$e^{\lambda \sigma^-}=-\lambda (x^-+p)$ (see e.g. ref. [\gidnel ]), while
in the second case the relevant coordinate  defining
$|0\rangle _{{\rm out},v}$ is $e^{\lambda \tau^-}=-\lambda (x^-+v)$,
as discussed in ref. [\yo ],
and $e^{\lambda \sigma^+}=e^{\lambda \tau ^+}=\lambda x^+$.
The false vacuum is thermally populated with respect to the true
vacuum and they are related by a Bogoluvob transform.

The issue of energy-conservation is subtle because it is
not clear what energy we should attribute to the thunderbolt.
The curvature being of Planckian order, our classical description
of gravitation and matter cannot be expected to remain valid
under these extreme conditions. In particular,
any detector that we can devise will be destroyed.
Inside the ``bubble" there should be  a combination of
negative energy density with positive energy corresponding to the
collapsed matter which will finally comes out carrying
the information back.
The ``Schwarzchild" observers who have calibrated their detectors with
the false vacuum, $|0\rangle _{{\rm out},p} $, i.e. by demanding
${}_{p,{\rm out}}\langle 0| T_{--}^{\rm (i)} |0\rangle _{{\rm out},p}=0\ ,
$
will measure the usual outgoing flux of Hawking radiation in
region I and receive a total energy of the same order
of the black hole ADM mass:
$$
{}_{\rm in}\langle 0| T_{--}^{\rm (i)} |0\rangle _{\rm in}=
{\kappa\over 4}\big[ {1\over (x^-+p)^2} - {1\over {x^-}^2} \big]\ ,
\ \ E^{(i)}=-\lambda \int _{-\infty}^{x^-_0} dx^- (x^-+p) T_{--}^{(i)}(x^-)
\cong m \ .
\eqn\tini
$$
Observers who measure energy with respect to the true final vacuum,
$|0\rangle _{{\rm out},v}$,
by demanding ${}_{v,{\rm out}}\langle 0| T_{--}^{\rm (i)}
|0\rangle _{{\rm out},v}=0 $,
will receive a very weak flux of
Hawking radiation in region I [\yo ],
$$
{}_{\rm in}\langle 0| T_{--}^{\rm (i)} |0\rangle _{\rm in}=
{\kappa\over 4} \big[ {1\over (x^-+v)^2} - {1\over {x^-}^2}
\big]\ .
\eqn\temenos
$$

The ingoing flux of negative energy explains why a presumptive
planckian-size object can engender such a violent event on an astronomical
scale.  The presence of negative energy is problematic;
 it threatens the stability of the vacuum, especially if
negative energy is reflected out affecting asymptotic observers.
Nonetheless, given the vacuum transition,
it is not clear that this ingoing flux
will necessarily convey negative energy into ${\cal J}^+$ after
the bouncing.

Now let us consider quantum gravity in 3+1 dimensions.
Any realistic quantum theory of gravity must have Einstein gravity as
classical limit. Our discussion in the previous section only entails
classical regions, and how to patch them together.
Therefore, in reproducing the previous analysis by using Einstein equations,
one would obtain a result valid for any theory of quantum gravity
in 3+1 dimensions which satisfies postulates 1 and 2.
Unfortunately, the general solution to Einstein equations coupled to matter
is unknown. Nevertheless, we hope that the mathematical complexity of the
3+1 dimensional theory can be somehow circumvented and, perhaps, a general
``quantum" singularity theorem is attainable.
Here we only discuss the general idea.
The radial null geodesics of the Schwarzchild metric satisfy
$t=\pm r_*+$const. , where $r_*$ is the tortoise coordinate defined by
$$
r_*=r+2M \log \big( {r\over 2M}-1\big)\ .
\eqn\tortoi
$$
In the Hawking's model of gravitational collapse [\hawi ],
the {\it in} modes are introduced by using  plane waves constructed
in terms of $r+t$, $r-t$, whereas
for the {\it out} modes the relevant coordinates are $r_*+t$, $r_*-t$.
The Kruskal coordinates $U,V$ are given by
$$
2M e^{r_*\over 2M}=(r-2M)e^{r\over 2M}=-V(U+2M)\ ,\ \ \
e^{t\over 2M}=-{V\over U+2M}\ .
\eqn\kal
$$
According to postulate 2, the Kruskal diagram of the time-dependent geometry
is as in fig. 3. This means that, instead of $r_*,t$,
the asymptotic coordinates for the {\it out} observers $R_*,T$ will have
to be of the form
$$
2Me^{R_*\over 2M}=-V(U+v)\ ,\ \ \
e^{T\over 2M}=-{V\over U+v}\ ,
\eqn\kal
$$
where $v<O(l_P)$.
For the reasons described in the previous section,
one expects that there should be a region of planckian curvatures
before entering into the vacuum region.
In fact, the assumption that the curvature is weak
in region II at ${\cal J}^+$ should again lead to a contradiction.
What we need is a  proof that
any classical solution in region II that is continuously matched
with the Schwarzchild geometry in region I
has either a curvature singularity or a boundary, at some $U\cong -2M $,
for any given $V>>V_e$ (the idea is that for classical
and continuous geometries the asymptotic
value of $U$ should be determined by the incoming matter;
it is only if there is a strong curvature region spreading out to
infinity that we can in principle have something different than $U\cong -2M $).
Now, the possibility of a boundary must be excluded by
postulate 2. On the other hand,
a discontinuity in the metric as a function of $U$ also begets
a curvature singularity at some $U\cong U_e$.
Postulate 2 requires that
this curvature singularity can somehow be continued through
due to quantum corrections.
Again one would be led to conclude that in a quantum theory of gravity
that satisfies postulates 1 and 2 a thunderbolt carrying
planckian curvature at $U\cong U_e $ is inevitable.

The arguments in these paper only hold for macroscopic black holes,
i.e., for black holes whose masses are much larger than the Planck masses,
where the semi-classical methods are applicable.
In particular, virtual, Planck-size black holes formed as
intermediate states in low-energy scattering processes may well evaporate
completely in a smooth way. Only large black holes will occasion a
violent event
at the endpoint in this scenario dictated by postulates 1 and 2,
but they cannot be virtual because Planck scale is a cut-off in
loop integrals.

To conclude,
we have seen in 1+1 dimensions that, if quantum gravity removes the
 black hole singularity leaving a time-like boundary,
an outgoing shock wave producing  planckian curvatures will emerge at the
end point of black hole evaporation. In addition,
semi-classical arguments suggest
that at the same time there must be an imploding
wave carrying macroscopic amounts of negative energy.
There is a  danger of negative energy being
reflected out, which arises some doubts about   vacuum
stability and thus about the consistency of the full picture.
The singularity problem, as well as the information problem, seem to be
resolved at a too high price. The scenario may nevertheless be viable, as
we were unable to find a clear inconsistency.
Assuming the present result holds true in 3+1 dimensions,
one could easily
establish whether there is a singularity inside an evaporating
black hole with no need of
going in: if we survive (e.g. if only a mild shock wave -a thunderpop-
is emitted at the endpoint of evaporation), then
topology change will have inexorably taken place.

There is a widespread belief that in superstring theory
singularities should be absent, since space-time loses its meaning
for distances less than the string size. One expectation is, in fact,
that gravitational  collapse in string theory should be described by a
Minkowski topology. Unfortunately, the arguments of this paper
would equally apply to this
case, and a catastrophe would likely occur at the moment the
evaporation is completed.

Irrespective of what is physics at Planck scales,
a formulation of quantum gravity in which only trivial topologies
are admissible is surprisingly beset by acute problems.

\bigskip
The author wishes to thank
A. Linde and L. Susskind for useful discussions.

\refout
\bigskip

\centerline {\bf Figure Captions}

\smallskip

\n {Fig. 1: Standard semi-classical
picture of black hole evaporation.}

\n {Fig. 2: Black hole evaporation described by a
topologically trivial Lorentzian metric.}

\n {Fig. 3: Kruskal diagram of black hole evaporation
in the scenario implied by postulates I and II.}

\vfill\eject
\end